\documentstyle[prb,aps]{revtex}

\begin{document}

\title{Interplay of Orbital Degeneracy and Superconductivity in a
Molecular Conductor}
\author{ Michele Fabrizio, Marco Airoldi and Erio Tosatti}
\address{
Istituto Nazionale di Fisica della Materia (INFM) and}
\address{
International School for Advanced Studies,
Via Beirut 4, 34014 Trieste, Italy}

\date{24 july 1995}
\maketitle
\begin{abstract}
We study electron propagation in a molecular lattice model. Each
molecular site involves doubly degenerate electronic states coupled
to doubly degenerate molecular vibration, leading to a so--called
E-e type of Jahn-Teller Hamiltonian.
For weak electron-phonon coupling and in the anti-adiabatic limit we
find that the orbital degeneracy induces an intersite pairing
mechanism which is absent in the standard non-degenerate polaronic model.
In this limit we analyse the model in the presence of an additional on-site
repulsion and we determine, within BCS mean field theory, the
region of stability of superconductivity. In one dimension, where powerful
analytical techniques are available, we are able to calculate
the phase diagram of the model both for weak and for strong electron-phonon
coupling.
\end{abstract}

\section{Introduction} \label{INTRODUCTION}

When atoms or molecules with orbitally degenerate valence levels are
arranged to form a solid,
the degeneracy of the isolated constituents is often broken, once the
solid is formed. In many cases, in fact, the crystal field, produced by
the surrounding atoms/molecules, is able to remove the original degeneracy.
Yet, if the crystal symmetry is sufficiently high
the orbital degeneracy may not be completely
lifted.  In this situation the Jahn-Teller (JT) effect, arising from
coupling electrons to vibronic modes, may play an important role. In
particular it can induce a global symmetry-lowering lattice
deformation lifting
the residual degeneracy (``static'' Jahn-Teller effect). However
if the phonon frequencies are high in comparison with the electron hopping,
the static distortion may become
disadvantageous and the original symmetry may be recovered
dynamically (``dynamic'' Jahn-Teller effect). A primary and well known
consequence of the dynamic mixing between electronic and vibronic degrees
of freedom, is the renormalization of several electronic
matrix elements by the so-called Ham factors.
Apart from this suppression factor, other interesting properties may
arise when dynamical JT effect is important.

For instance, in the context of superconductivity in fullerenes,
it has been recently proposed\cite{Erio} that
the dynamical JT effect may be associated with an increase of the
electronic pairing interaction. In the specific case
of charged fullerene molecules, the JT effect
arises mainly from coupling the partially occupied $t_{1u}$ orbitals with the
$H_g$ vibronic modes, even though for a realistic description one has to
take into account other modes such as $A_g$ \cite{Gunnarsson,Eriobis}
and other orbitals, such as $h_u$ and $t_{1g}$. Since the physical model is
very complicated,
in Ref.\onlinecite{Manini} a simplified version was introduced,
aimed at capturing
the essential physics of the problem. The model
consists of a lattice of molecules, each with two degenerate orbitals coupled
to a doubly-degenerate vibronic mode. The Hamiltonian of each molecule is
described by a so-called E-e JT Hamiltonian, which is the simpler
case of dynamic JT effect\cite{Englman}.
In spite of its simplicity,
the lattice of E-e molecules was shown\cite{Manini,Maninibis} to exhibit
in the strong coupling limit rather striking and unexpected features.
In particular it was found that, even if the polaronic attraction is
disregarded, two electrons in the vacuum still bind with a binding energy
proportional
to the effective hopping (the bare hopping reduced by the Ham factor).
At higher density however the model could only be analyzed numerically and
in one dimension\cite{Maninibis}.

Besides the application to $C_{60}$, the doubly
degenerate Jahn-Teller model might provide useful information for other
systems where the Jahn-Teller effect is expected to be important,
such as compounds containing magnetic ions with unfilled $d$ or $f$ shells.
In this case it is well known that the interplay between the
Jahn-Teller effect and the strong electronic correlations plays a very
important role in determining both the structural and the magnetic
properties (for a review mainly on transition metal compounds see e.g.
Ref.\onlinecite{Englman} and Ref.\onlinecite{Khomskii},
and for rare earth compounds see Ref.\onlinecite{Gehring}).
A typical example where the doubly degenerate JT model could be
relevant is a transition metal ion, whose valence state
is five-fold orbitally degenerate.
In cubic symmetry, the crystal field splits the $d$-levels
into a three-fold degenerate $t_{2g}$ level and a two-fold degenerate
$e_g$. Exactly for a tetrahedral distortion, the
two-fold degenerate $e_g$ is the ground state.

In this paper we study the lattice of E-e molecules in the
weak electron-phonon coupling regime and for phonon frequencies
higher or comparable to the electron bandwidth.
Since any perturbation,  however small,
has always important consequences for degenerate levels,
most of the interesting features originally
recognized in Ref.\onlinecite{Maninibis} for strong electron-phonon coupling,
are already present in the weak coupling regime. This limit has in addition
the big advantage of allowing an analytical approach which will
provide new useful results valid in any dimension.
Moreover in one-dimension, where powerful analytical techniques are available,
we are able to characterize the whole phase diagram of the model,
both in the weak and in the strong coupling regimes.

The paper is organized as follows. In Section \ref{MOLECULE} we introduce
the model and discuss the properties of a single molecule both in the weak
and strong coupling limit. In the former limit, we find an unitary
transformation which, when applied to the molecular Hamiltonian,
provides an effective Hamiltonian without electron-phonon coupling which
nevertheless can reproduce all the correct results up to fifth order
in the electron-phonon coupling constant.
A lattice of such JT molecules is introduced and analyzed in
Section \ref{LATTICE}.
In Section \ref{BCS} we study the model within a
BCS mean field approach. Finally, in Section \ref{ONE_D}, we calculate
the phase diagram of the model in one-dimension.

\section{The molecule} \label{MOLECULE}

The model we are going to discuss describes an array of molecules
with two degenerate electronic orbitals $c_{1\sigma}$ and $c_{2\sigma}$
(later referred to as ``band''),
coupled to a two-dimensional molecular vibration (henceforth
called phonon) with energy $ \omega_0$ ($\hbar=1$). Each molecule is
described by the so called E-e Hamiltonian:
\begin{equation} \label{MODEL:eqn}
H_{mol} = \frac{\omega_0}{2}
\left( \vec{r}\,^2 + \vec{p}\,^2 \right) + g \omega_0
\vec{r}\cdot\vec{\tau},
 \end{equation}
where $\vec{r}=(x,y)$ is the two-dimensional coordinate of the local
phonon mode, and
\begin{equation}\label{tau}
\vec{\tau}=\frac{1}{2} \sum_{a,b=1,2} \sum_{\alpha=\uparrow,\downarrow}
c^\dagger_{a\alpha}\vec{\sigma}_{ab}
c^{\phantom{\dagger}}_{b\alpha},
\end{equation}
being $\vec{\sigma}$ the Pauli matrices.

The Hamiltonian (\ref{MODEL:eqn}) has to be compared with that
describing single-band electrons coupled to a non-degenerate phonon
\begin{equation}\label{standard}
H_{mol}=\frac{\omega_0}{2}
\left( r^2 + p^2 \right) + g \omega_0 r n,
\end{equation}
where $r$ is the one-component phonon coordinate and
\[
n= \sum_{\alpha=\uparrow,\downarrow}
c^\dagger_{\alpha}c^{\phantom{\dagger}}_{\alpha},
\]
is the local density of single band electrons.
In what follows we will show that the additional degrees of freedom
of (\ref{MODEL:eqn}) give rise to new interesting properties.

In the absence of electron-phonon coupling, each molecular level
of the Hamiltonian (\ref{MODEL:eqn}) with vibronic energy
$m \omega_0$ and $n$ electrons is degenerate, with degeneracy
\[
(m+1)\times \left( \begin{array}{c} 4 \\ n \\ \end{array} \right).
\]
The binomial coefficient counts the number of ways of distributing
$n$ fermions
among 4 levels (1 and 2, $\uparrow$ and $\downarrow$). These states can be
for instance labeled by the total spin $S$, its {\sl z}-component $S_z$,
and by $\tau_z$ (which is half the difference between the number of electrons
in orbital 1 and that one in orbital 2). The factor (m+1) is instead the
degeneracy of
each vibrational state and corresponds to the possible values that the
vibron angular momentum $L_z=xp_y - yp_x$ can
assume ($L_z=-m,-m+2,...,m$).
When the electron-phonon coupling is switched on, this degeneracy is
lifted. The total spin and its {\sl z}-component are
still good quantum numbers,
but now only the {\sl z}-component of the total pseudo-angular momentum
\[
J_z=L_z + \tau_z
\]
commutes with the Hamiltonian\cite{Englman}. Notice that
for odd number of electrons $\tau_z$ is half an odd integer, and
consequently is $J_z$. Due to the symmetry $J_z\to -J_z$,
each state (also the ground state) with odd number of electrons is
at least four-fold
degenerate ($\pm J_z$, $S=\pm 1/2$). On the contrary for even number of
electrons the starting degeneracy is split and the ground state
turns out to be an orbital as well as a spin singlet.

At weak electron-phonon coupling $g\ll 1$ each multiplet is split by
energy shifts of order $\omega_0 g^2$, but different
multiplets are still well separated by energy $\omega_0$.
If we are interested in the behavior of the model at energies
$\ll  \omega_0$, we can neglect all but the lowest multiplet. Since this
is adiabatically connected to the multiplet without excited vibrons
($m=0$, $L_z=0$), it can be labeled by the electronic quantum numbers only
($S$, $S_z$ and $\tau_z$). This suggests that it is possible to define
an effective Hamiltonian for this lowest multiplet which acts only on the
electronic degrees of freedom and is able to reproduce the energy shifts
inside the multiplet. A standard way
to derive this Hamiltonian is via an unitary transformation $U=e^S$.
For the non-degenerate model (\ref{standard}), this unitary transformation
simply shifts the origin of the harmonic oscillator to a new one, namely
$S=-igpn$. In the degenerate two-band model, the components of the new origin
$(g\tau_x,g\tau_y)$ do not commute among themselves, and this gives rise to
some of the interesting properties of the model.
Up to order $g^3$, the operator $U$ for the degenerate model is given by
(see e.g. M. Wagner in Ref.\onlinecite{Wagner} for a comprehensive review of
unitary transformations in JT models)
\begin{equation}\label{UnitaryTrasf:eqn}
U=e^{ -i g \vec p \cdot \vec \tau +i {g^3 \over 3} [(\vec x \cdot \vec x)
(\vec p \cdot \vec \tau) - (\vec x \cdot \vec \tau)
(\vec x \cdot \vec p)] +i g^3 \vec x \cdot (\vec \tau_z \times \vec
\tau)}.
\end{equation}
The transformed molecular Hamiltonian reads
\begin{eqnarray}
& &U H_{mol} U^{-1} = \frac{ \omega_0}{2}
\left( \vec{r}\,^2 + \vec{p}\,^2 \right) -
\omega_0 \frac{g^2}{2}\left(1-\frac{g^2}{4}\right) L_z \tau_z\nonumber\\
& & - \omega_0 {1 \over 2} g^2 (1-{g^2 \over 2}) \vec \tau\,^2
- \omega_0 {3 \over 8} g^4 \tau^2_z + \Delta H,
\label{EffHamiltonian:eqn}
\end{eqnarray}
where $\Delta H$ contains terms coupling the phonon modes with the electrons
and having coupling constants of order $O(g^4)$\cite{nota1}.
If we are interested in
physical quantities with a precision up to $g^5$, we can neglect $\Delta H$.
The molecular ground state with an odd number of electrons ($n=1,3$) is
four--fold degenerate ($J_z=\pm 1/2,S_z=\pm 1/2$), and
its ground state energy is,
up to order $g^4$, $E_{n=1,3}/\omega_0=-g^2/4 + g^4/32 $.
On the other hand, for two electrons, the electron-phonon
interaction splits the initial six--fold degenerate ground state state into a
multiplet whose lowest member is a non--degenerate singlet ($J_z=0,S=0$).
More precisely (in units of $\omega_0$):
\begin{equation} \label{ENERGY2:eqn}
\begin{array}{lcc}
\displaystyle
E_2(J_z=0,S=0) &=& -g^2 + g^4/2, \\
\displaystyle
E_2(J_z=\pm 1,S=0) &=& -g^2/2 - g^4/8,  \\
\displaystyle
E_2(J_z=0,S=1) &=&  0. \\
\end{array}
\end{equation}

In the strong coupling limit $g\gg 1$, the situation is quite different.
The ground state has the same quantum numbers as in the weak coupling limit,
i.e. $S=1/2$ and $J_z=\pm 1/2$ for odd numbers of electrons, and $S=0$ and
$J_z=0$ for even numbers. On the contrary the lowest excited states are
identified by the same spin of the ground state but higher $J_z$
(apart from the trivial case of 0 and 4
electrons, where there is no Jahn-Teller distortion). They are separated
from the ground state by an energy of the order $\omega_0 J_z^2/g^2$.
In order to describe this limit, Ref.\onlinecite{Manini}
introduced an effective model with a single electronic level
which guarantees that the doubly occupied site is always in a singlet state.
The larger occupancies were disregarded by imagining a strong on-site
repulsion able to cancel the strong polaronic binding energy (of order
$\omega_0 g^2$). The role of the quantum number $J_z$ was played by a quantum
rotator. The main difficulty with this representation is to implement
the constraint that $J_z$ has to be half integer for a singly occupied
site and integer otherwise. This condition is automatically verified
in the original two-band model, but has to be enforced by imposing a
constraint once the effective single-band plus pseudo-spin picture is
used.

\section{Lattice of molecules} \label{LATTICE}

Let us consider a lattice of E-e molecules coupled by the single particle
hopping term:
\begin{equation}
\displaystyle
H_{hopping} = - t \, \sum_{<i,j>} \sum_{\sigma}
\; \big [ c^{\dagger}_{1,i,\sigma} c_{1,j,\sigma} +
c^{\dagger}_{2,i,\sigma} c_{2,j,\sigma} \big ].
\label{bare-hopping}
\end{equation}
Electron hopping between two neighboring molecules modifies both
their spin $S$ and their pseudo angular momentum $J_z$. Therefore it will mix
the ground state configurations of each molecule with the excited states
of the others.
In the weak coupling limit $g\ll 1$, if moreover $tg\ll \omega_0$, we can
retain just the hopping processes which mix the states in the lowest
multiplet for each electron occupancy. In fact by means of the unitary
transformation Eq.(\ref{UnitaryTrasf:eqn}), we find:
\begin{eqnarray}
\displaystyle
H_{hopping} &=& - t \, \sum_{<i,j>} \sum_{\sigma}
\; U\big [ c^{\dagger}_{1,i,\sigma}
c^{\phantom{\dagger}}_{1,j,\sigma} +
c^{\dagger}_{2,i,\sigma} c^{\phantom{\dagger}}_{2,j,\sigma}
\big ]U^{-1}=\nonumber \\
\displaystyle
& &- t'  \, \sum_{<i,j>} \sum_{\sigma}
\left\{ c^{\dagger}_{1,i,\sigma} c^{\phantom{\dagger}}_{1,j,\sigma}
\left[ 1 + \frac{g^4}{8}\left(n_{2i-\sigma}+n_{2j-\sigma}
- n_{1i-\sigma} - n_{1j-\sigma}\right)\right]\right. \label{hopping}\\
\displaystyle
& &  -  \frac{g^4}{8}c^{\dagger}_{1,i,\sigma}
c^{\phantom{\dagger}}_{1,j,-\sigma}
\left(c^{\dagger}_{2,i,-\sigma} c^{\phantom{\dagger}}_{2,i,\sigma}+
c^{\dagger}_{2,j,-\sigma} c^{\phantom{\dagger}}_{2,j,\sigma}\right)
\nonumber \\
& & \left. -\frac{g^4}{2}\left[
c^\dagger_{1,i,\sigma}\tau^-_j c^{\phantom{\dagger}}_{2,j,\sigma}
\tau_{z,j} + \tau_{z,i} c^\dagger_{2,i,\sigma}\tau^+_i
c^{\phantom{\dagger}}_{1,j,\sigma}\right]
+ \left(1\leftrightarrow 2\right) \right\}, \nonumber
\end{eqnarray}
where $t'=t(1-g^2/8 + 3g^4/64)$, and this effective hopping is intended
to give the correct results for $tg\ll \omega_0$.
The last term in square brackets acts only when a triply occupied site
is involved in the hopping process, and is crucial to maintain the
particle-hole symmetry around half-filling (two electrons per site).
The effective
hopping Hamiltonian (\ref{hopping}) plus the molecular
term (\ref{EffHamiltonian:eqn}) therefore describe the lattice of E--e
molecules for sufficiently small electron-phonon coupling $g$,
even in the interesting weakly anti-adiabatic limit $t \sim \omega_0$.

{}From Eq.(\ref{hopping}) we see that
the electron--phonon interaction modifies the hopping
amplitude according to the occupation of the sites involved in the hopping
process. In particular, if we restrict to the lowest-energy molecular states,
the hopping amplitude from (or into)
a doubly occupied site ($J_z=S=0$) increases relatively to the hopping
from a single occupied site (also in the molecular ground state)
to an empty one. For instance the hopping process from a
doubly occupied site to a nearest neighboring empty site, relatively to
that from a singly occupied site to an empty site is
$T_{2\to 0}/T_{1\to 0} = (1+g^4/4)/\sqrt{2}$ for small $g$
(see Fig.~1), while
$T_{2\to 0}/T_{1\to 0} \to 1$ at large $g$\cite{Manini}.
As soon as this ratio $T_{2\to 0}/T_{1\to 0}>1/\sqrt{2}$ (i.e. $g\not = 0$)
a two particle bound state appears in one and two dimensions, even if
we neglect the polaronic
binding energy. For instance the two-particle problem in vacuum is
easily solved (in analogy with Ref.\onlinecite{Maninibis}) and
the self--consistency condition for the energy $E$ reads:
\begin{equation} \label{E:eqn}
\frac{1}{N} \sum_{k} \frac{1}{E- 2 \epsilon_k} =
\frac{ 1 + g^4/2 }
{ E g^4/2 },
\end{equation}
where $\epsilon_k$ is the hopping energy in momentum space. This equation
indeed admits a bound state solution $E<2\epsilon_0$ in one and
two dimensions as soon as
$g\not = 0$. Notice that this bound state is a feature peculiar to the
degenerate model and it is absent for the non-degenerate version
(\ref{standard}).

At strong coupling the situation is more complicated. In fact the number
of lowest excited states, characterized by higher $J_z$,  with excitation
energy $\leq \omega_0$, grows like
$g$ for large $g$, and therefore greatly exceed the analogous number
in the weak coupling limit (which coincides with the number of states in the
lowest multiplet, i.e. six).
In order to simplify the analysis, in Refs.\onlinecite{Manini,Maninibis}
the excitations into these higher-$J_z$ states were forbidden. This amounts
to assume that the matrix elements due to the hopping processes
which mix these states
with the molecular ground state configurations, are much smaller than the
excitation energies, that is
\[
t\ll \frac{\omega_0}{g^2},
\]
which is therefore the limit of validity of the results found in
Ref.\onlinecite{Maninibis}. Even with this simplification, the model in the
strong coupling limit remains analytically quite intractable due to
the constraint that $J_z$ should be half integer for odd occupancy and
integer otherwise, as previously discussed. This is the reason why the
analysis\cite{Maninibis} was done only numerically and in one-dimension.

\section{BCS-Mean Field Solution.}\label{BCS}

In the previous Sections we have shown how the lattice of E-e molecules
can be mapped in the weak coupling limit ($g\ll 1$ and $tg\ll \omega_0$)
onto the model with the Hamiltonian (\ref{EffHamiltonian:eqn}) plus
(\ref{hopping}) where only electronic degrees of freedom appear. This
model can easily be analyzed by standard many-body techniques. For instance
we can study within BCS mean-field theory the instability to
superconductivity.
In order to describe a more realistic system, we also include
a generalized on-site interaction including Hund's rule exchange in the form:
\begin{equation}\label{onsiterepulsion}
\displaystyle
{U \over 2} \sum_{\sigma}(n_{1 \sigma}n_{1 -\sigma} + n_{2 \sigma}n_{2
-\sigma})
+ V\sum_{\sigma\sigma'} n_{1\sigma}n_{2\sigma'} -
\Gamma \vec{S}_1\cdot\vec{S}_2,
\end{equation}
where $\vec{S}_a$ is the spin operator of electrons $a=1,2$. $U$, $V$ and
$\Gamma$ is in fact the minimal set of parameters describing this six--state
three--level multiplet of the doubly occupied site.

The BCS wave function we use to minimize the energy is
\[
\displaystyle
|\Phi_0\rangle = \prod_k \left[
u_k + \frac{v_k}{\sqrt{2}}\left( c^\dagger_{1k\uparrow}
c^\dagger_{2-k\downarrow}  +  c^\dagger_{2-k\uparrow}
c^\dagger_{1k\downarrow}\right)\right] |0\rangle \,.
\]
The interaction between the Cooper pairs is given by:
\begin{eqnarray*}
V_{{\bf k k}'} &=& V +\frac{3}{4}\Gamma - \frac{1}{2} \omega_0 g^2 +
\frac{7}{16} \omega_0 g^4 \\
& &-t g^4\sum_{i=1,d}\left[\cos(k_i a) + \cos(k'_i a)\right],
\end{eqnarray*}
where {\bf k}$=(k_1,k_2,...,k_d)$ is the relative
momentum of the pair. The BCS equations are:
\begin{eqnarray*}
\Delta_{{\bf k}} &=& - {1 \over L^D} \, \sum_{{\bf k}'} V_{{\bf k k}'} \,
\frac{ \Delta_{{\bf k}'}}{2 E_{{\bf k}'} }\, ,  \\
E_{{\bf k}} &=& \sqrt {(\epsilon_{{\bf k}} -\mu)^2 + \Delta_{\bf k}^2 }\, , \\
\displaystyle
\epsilon_{{\bf k}} &=& -2t'
\sum_{i} \cos (k_i a)\, , \\
n &=& 1 - {1 \over L^D} \sum_{{\bf k}} \left( {\epsilon_{\bf k}
-\mu  \over E_{\bf k}}
\right) \, ,
\end{eqnarray*}
being $n$ the electron density and $\mu$ the chemical potential.
The form of the BCS equations implies that the gap $\Delta_{\bf k}$
depends on the wavevector {\bf k} only through
the free particle
dispersion $\epsilon_k$ and this dependence is linear. Therefore
$\Delta_{\bf k}$ can be parametrized by the two unknowns $\Delta$ and $\chi$:

\begin{equation}
\Delta_{\bf k} = \Delta \, [1+\chi \sum_{i} \cos( k_i a)],
\end{equation}
and the BCS set of equations reduce to a set of coupled equations for $\Delta$,
$\chi$ and the chemical potential $\mu$ which must be solved numerically.
The critical line between the superconducting and the normal state
can be obtained analytically:
\begin{equation}\label{ucritbcs}
V_c+\frac{3}{4}\Gamma_c = \frac{1}{2}\omega_0 g^2
-\frac{7}{16}\omega_0 g^4
- g^4 \mu  + O(g^6).
\end{equation}
The first two terms on the right hand side represent a negative pairing
energy, originating from a gain of molecular zero--point energy upon
pairing~\cite{Erio}.
Finally the correlated hopping contribution $g^4 \mu$ provides an
additional pairing mechanism (note that $\mu < 0$) which is the analog
of that described in Ref.\onlinecite{Maninibis} in the
electron--pseudospin model.
This term, being intersite, is favored by a large coordination
number. We recall that this term is originated by the degeneracy of the
electronic band and vibronic modes and has no equivalent in a standard
non degenerate polaronic model.

\section{One-dimensional phase diagram} \label{ONE_D}

It is interesting to compare the properties of the degenerate two-band model
with those of a single-band model in one-dimension (1D), where
several rigorous results are known for both weak and strong
electron-phonon coupling. In the following analysis we will restrict
to densities less or equal to one electron per site.

A peculiar feature of 1D systems, which is absent in higher
dimensions but for particular band structures and fillings, is
the nesting property of the Fermi surface. This induces a strong coupling
between the $2k_F$-phonons and the $2k_F$ Charge Density Waves (CDW) (Peierls's
instability). In fact, as a result of the diverging $2k_F$ density-density
electronic correlation function, the
$2k_F$ phonon frequency softens which in turns
leads to an increase of the coupling of these phonons with
the CDW. The lower the bare $2k_F$-phonon
frequency compared to the electron bandwidth, or alternatively the stronger
the electron-phonon coupling constant,
the more favored the CDW with respect to Singlet
Superconductivity (SS) (see e.g. Ref.\onlinecite{Bourbonnais}).

\subsection{Non degenerate single--band model}
In the standard single-band model (\ref{standard}),
the resulting phase diagram has a region at large phonon frequency and small
electron-phonon coupling where SS dominates. Upon decreasing the $2k_F$-phonon
frequency or increasing the electron-phonon coupling constant,
the model has a cross-over to a region where CDW dominates.
Further inside this region a better approach is to represent the
$2k_F$ lattice-distortion with its phase and amplitude. The amplitude
fluctuations around the {\sl finite} average value can be neglected.
On the contrary the phase fluctuations are gapless and so strongly coupled to
the CDW to be practically indistinguishable
(the two move together) (see e.g. Ref.\onlinecite{Fukuyama}).
Although the two regimes look different (at weak coupling the amplitude
of the lattice distortion is still not developed due to quantum fluctuations)
there is no real transition between them and only a cross-over from
SS to CDW characterizes the overall phase diagram, as qualitatively
drawn in Fig.~2. An additional electron-electron repulsion enlarges
the regime of dominant CDW at expenses of SS\cite{Bourbonnais,Fukuyama,Voit}.

\subsection{Degenerate two-band model: weak coupling limit}
We start our analysis of the two-band model from the limit of
weak electron-phonon coupling and high phonon frequency.

\noindent
{\bf a)} \, {\it Two-band model, no electron-electron interaction.}
If we integrate out the phonon field we
obtain the effective retarded electron-electron interaction
\begin{equation}
- \omega_0\left( \frac{g^2}{2}\right)\,
\frac{\omega_0^2}{\omega_0^2-\omega^2}
\vec{\tau}(\omega)\cdot \vec{\tau}(-\omega).
\label{eff-interaction}
\end{equation}
If $g\ll 1$ and $\omega_0 \gg t$, the effective interaction
can be approximated by its unretarded limit
\[
- \omega_0 \frac{g^2}{2}
\vec{\tau}(\omega)\cdot \vec{\tau}(-\omega).
\]
This route has been recently followed by Shelton and Tsvelik\cite{Tsvelik}
who treat it further
by means of bosonization plus Renormalization Group (RG) analysis,
in the absence of electron-electron interaction.
In the bosonization language, the fermion model
is described by sound modes. In the present case one can define four of
these modes:
symmetric (with respect to the band indices 1 and 2) and anti-symmetric
charge and spin sound modes. Shelton and Tsvelik have found that all these
modes acquire a gap except for the symmetric charge mode. This
remains gapless
and is identified by a Luttinger liquid exponent
$K={\rm e}^{2\varphi}<1$\cite{Haldane}. The larger the
phonon-induced electron-electron interaction $g^2 \omega_0$, the smaller
$K$.
For $K>1/2$ the dominant fluctuations are indeed the SS ones, with the
pairing operator we have introduced in the previous section. For
$K<1/2$, on the contrary, $4k_F$-CDW dominate. As before, one finds a smooth
cross-over from SS at weak coupling to CDW at strong coupling.

\noindent
{\bf b)} \, {\it Two-band model with finite electron-electron interaction.}
We consider the model in the same weak coupling limit
as in Ref.\onlinecite{Tsvelik} but in the presence of the
interaction Eq.~(\ref{onsiterepulsion}). It is useful to begin showing how
the interaction modifies the energies of the multiplet
(\ref{ENERGY2:eqn}), since the ordering of these levels determines the
physical behavior of the model:
\begin{eqnarray}
E_2(J_z=0,S=0) &=& -\omega_0 g^2 + \omega_0 \frac{g^4}{2} +
V+\frac{3}{4}\Gamma ,\nonumber\\
E_2(J_z=\pm 1,S=0) &=& -\omega_0 \frac{g^2}{2} - \omega_0 \frac{g^4}{8}
+ U,  \label{multiplet}\\
\displaystyle
E_2(J_z=0,S=1) &=&  V -\frac{1}{4}\Gamma. \nonumber
\end{eqnarray}
We see that the $J_z=S=0$ state remains the lowest energy state
until the Hund term $\Gamma$ is so large to make the triplet
state favorable or until the inter-orbital Coulomb repulsion $V$
does not exceed the intra-orbital $U$ so much to make the
$J_z=1$ state favorable. In what follows this latter possibility
will be disregarded, i.e. we will always assume $U\geq V$.

By including the interaction within the RG
approach of Ref.\onlinecite{Tsvelik}, we find that,
so long as the $J_z=S=0$ state remains the
lowest energy state, the physical behavior of the model
does not change qualitatively with respect to the case in the
absence of interaction. The main effect of the interaction is to
diminish the Luttinger liquid exponent $K$, which favors CDW against SS.
In addition, as the on-site Coulomb repulsions
$U$ and $V$ increase, the $4k_F$-CDW (where the charge fluctuates from
0 to 2 electrons) is suppressed in favor of the $8k_F$-CDW
(where the fluctuation is between charge 0 and 1).

Analogous results were first obtained by Manini {\it et al.}\cite{Maninibis},
by numerical diagonalization of the effective model at $g\gg 1$
and $\omega_0 \gg tg^2$. They also analyze the model with an
additional repulsive interaction [practically their case corresponds
to $U=2V$ and $\Gamma=0$ in Eq.~(\ref{onsiterepulsion})], part of which
cancels the strong polaronic attraction (of order $\omega_0 g^2$).
Thus the only pairing mechanism they are left with
is that induced by the correlated hopping corresponding to our
$g^4 \mu$, as previously
discussed. On the contrary Shelton and Tsvelik only consider the effects
of the polaronic attraction, which is dominant at small $g$.
The two apparently different approaches give analogous results since,
as we have shown, correlated hopping favors a pairing
in the same channel as the polaronic term (i.e. both orbital
and spin singlet).

\noindent
{\bf c)} \, {\it Two-band model, no electron-phonon coupling.}
The model with the interaction term (\ref{onsiterepulsion})
but no electron-phonon coupling has also interesting properties.
If $\Gamma\not =0$ we still find gaps in the antisymmetric charge
sector and in the spin sectors. Moreover, if $K<1/2$, we find in addition
dominant singlet superconductive fluctuations, which however differ from
those we find in the presence of electron-phonon coupling since they are
odd by interchanging the two orbitals (which is a way of avoiding the on site
repulsion). This curious result which predicts superconductivity for
purely repulsive interaction is indeed a common feature of many two-band
models in one dimension\cite{twochains}.
For $K>1/2$ the $4k_F$ CDW fluctuations become dominant, although
the pair correlation function is still power-law decaying.
In the presence of phonon induced
electron-electron interaction (which we still take as unretarded),
this phase remains stable as far as the doubly occupied site
in a triplet configuration is lower in energy than the singlet one;
otherwise, as we already said, the previously discussed regime occur.
The two phases are separated by a gapless Luttinger liquid
regime with interaction--dependent exponents.

\noindent
{\bf d)} \, {\it Quarter-filled two-band model}.
At quarter filling (one electron per site) $8k_F$-Umklapp scattering is
possible and, if $K<1/2$, the symmetric charge mode gets gapped and
the system becomes insulating. This occurs
if either the electron-phonon coupling or the on-site repulsion are
sufficiently strong. In this commensurate phase all modes are gapped
unless the coupling constants have particular values. For instance
if the electron-electron interaction and the electron-phonon coupling combine
in such a way that the doubly occupied
site multiplet (\ref{multiplet}) is ``accidentally''
degenerate, then the antisymmetric charge mode and the spin modes
are gapless. At weak coupling we in fact realize that all the scattering
processes, apart from the Umklapp, are marginally irrelevant.

For strong repulsion, the model can be mapped onto a generalized Heisenberg
chain with two species of spin-1/2 per site: the physical spin $\vec{S}$ and
the electron pseudo angular momentum $\vec{\tau}$ defined by
Eq.(\ref{tau}).
This kind of super-exchange Hamiltonians have been quite intensively
studied in the context of insulators containing transition metal
ions (see e.g. Ref.\onlinecite{Khomskii}). Practically the derivation of
the effective Hamiltonian follows the derivation of the Heisenberg
model from the Hubbard model at half filling for $U\gg t$.
The important difference is that the state of a virtually doubly-occupied site
is not unique. The two electron state can be any of the levels
in the multiplet (\ref{multiplet}). Accordingly we can define three
super-exchange couplings: $J_{sing}$ if the intermediate doubly occupied level
has $S=J_z=0$, $J_T$ if it has $S=1$ and $J_z=0$, and $J_1$ if
$S=0$ and $J_z=\pm 1$. The effective 1D Hamiltonian is:
\begin{eqnarray}
H_{eff}&=& \left(\frac{J_{sing}}{2} - \frac{J_T}{2} + J_1 \right)
 \sum_{<ij>} \vec{S}_i\cdot \vec{S}_j \label{HHeis}\\
&+& \left( \frac{3J_T}{2} - \frac{J_{sing}}{2} \right)
 \sum_{<ij>} \vec{\tau}_i \cdot \vec{\tau}_j
 + \left(J_{sing} - J_1 \right)\sum_{<ij>} \tau_{z i}\tau_{z j}
\nonumber\\
&+&\left( 2J_T + 2J_{sing} \right) \sum_{<ij>}
\left(\vec{S}_i\cdot \vec{S}_j \right)\,
\left( \vec{\tau}_i \cdot \vec{\tau}_j\right)
+\left( 4J_1 - 4J_{sing}\right) \sum_{<ij>}
\left(\vec{S}_i\cdot \vec{S}_j\right)  \,
\left(\tau_{z i}\tau_{z j}\right)\nonumber.
\end{eqnarray}
Similar Hamiltonians
have been recently introduced also in connection with fullerides.
For instance, if the polaronic attraction is so strong that the
$J_z=S=0$ state for two electrons is much lower in energy than the other
levels, then one can assume $J_T=J_1=0$. In this case it has been shown
in Ref.\onlinecite{Maninibis} that
the spectrum of $H_{eff}$ has a gap both in the spin and in the
pseudo spin sector and the ground state is a kind of valence bond
solid. If, on the contrary, the triplet state energy is lower than the
singlet one,  the physical spins are ferromagnetically correlated.
In this case the Hamiltonian (\ref{HHeis}) has been recently invoked
by Auerbach {\it at al.}\cite{Auerbach} to explain the weak
ferromagnetism plus insulating behavior in TDAE-C$_{60}$.
These authors also identify some solvable points in the
phase diagram. For instance if $J_{sing}=J_T=J_1$ the model is
SU(4) invariant and is a particular case of a wide class of SU(N) invariant
models solved by Sutherland\cite{Sutherland}. This point corresponds to
the situation discussed above where the levels of the two-electron
multiplet become degenerate. In this case we indeed find
at weak coupling a spectrum with three gapless
excitations (the antisymmetric charge mode and the two spin modes)
which is in agreement
with the exact solution of the effective model at strong
coupling\cite{Sutherland}.

\subsection{Degenerate two-band model: strong coupling limit}
We now consider the limit of low phonon frequency or, equivalently,
of strong electron-phonon coupling.

\noindent
{\bf (i)} \, {\it Weak retardation effects.} We can improve the weak
coupling analysis
by taking into account a weak retardation.
The main consequence of a finite phonon-frequency
is its softening induced by the
coupling with the CDW. This in turn leads to an increase of the
electrons $2k_F$-phonon coupling constant and therefore to a decrease
of $K$. A simple method to describe this effect in the weak coupling
or high phonon frequency limit is the
two cut-off renormalization group (see e.g. Ref.\onlinecite{Bourbonnais}).
By applying this method we find that, as in the standard non degenerate model,
a decreasing phonon frequency is equivalent to an increasing coupling
constant.

\noindent
{\bf (ii)} \, {\it Strong retardation effects.}
New features appear as soon as the phonon frequency gets
too small or the electron-phonon coupling too large. In this
case we can no longer apply the weak coupling renormalization group, and we
have to resort to some strong coupling approach.
However we can qualitatively predict, by simple arguments, what is going to
happen in this limit.  For an isolated molecule we already saw that the
lowest excited states
in the strong coupling limit have the same total spin as the ground state
(e.g. $S=0$ for two
electrons) but higher $J_z$; their excitation energies vanish at large
$g$ approximately like $\omega_0 J_z^2/g^2$. When the molecules are
coupled, the hopping $t$, if not too small in comparison
with $\omega_0 /g^2$, will
clearly mix in the ground state wave function
molecular states with higher $J_z$. The $z$-component of the total
pseudo-angular momentum ${\cal J}_{z}$, i.e. the sum over all molecules
of $J_z$, is anyway a constant of motion,
and remains zero in the ground state. However we expect that
an excitation which changes the total pseudo-angular momentum
should become gapless in this strong coupling limit.
Notice that this regime falls outside that analyzed in
Ref.\onlinecite{Maninibis}.
Even though these authors consider the $g\gg 1$ limit, they still assume
$t\ll \omega_0 /g^2$ so that higher ${\cal J}_{z}$ states are always
separated by a finite gap in their case.

The previous qualitatively scenario can be formally derived by the
same approach which is used in the strong coupling limit for a
non-degenerate model (see e.g. Ref.\onlinecite{Fukuyama}).
The method consists in solving the mean field equations
of the 1D charge and spin sound waves (which are the only coherent excitations
in one dimension) strongly coupled to the phonon modes and then
expanding around the solution to take into account quantum fluctuations.
In our case we have to describe the $2k_F$ lattice distortion by
the amplitudes $\Delta_x=\Delta\cos(\theta)$ and
$\Delta_y=\Delta\sin(\theta)$ and the phases of each phonon component
($x$ or $y$). The two phases are locked to the symmetric charge mode
and the three describe a single gapless sound mode. Moreover $\theta$ gets
locked to the antisymmetric charge mode and the two also describe
a single gapless mode. The amplitude $\Delta$, which acquires at mean field
level a finite value, feels on the contrary
a finite restoring force, so that its fluctuations can be neglected at low
temperature. Analogously, the spin modes get gapped.
In conclusion we find that: 1) the gap of the anti-symmetric
charge mode vanishes while the spin gaps remain finite; 2)
the dominant fluctuations are now inter-band
charge density waves or better pseudo angular momentum density waves
identified by the $2k_F$ components of the operators
\[
\tau^+ = c^\dagger_{1\sigma}c^{\phantom{\dagger}}_{2\sigma}\;\; ,\;\;
\tau^- = c^\dagger_{2\sigma}c^{\phantom{\dagger}}_{1\sigma}.
\]
In a sense we find in this regime an orbitally quasi-ordered ground state
(in 1D a continuous symmetry can not be broken).

The vanishing gap in the antisymmetric charge mode can also
be inferred from the weak coupling regime. In fact as the electron-phonon
coupling increases we find that the operator
which couples the charge mode to the spin modes gets more and more
irrelevant, so that even if the spin modes remain gapped they will eventually
become unable to induce a charge gap. Therefore the two approaches,
weak and strong coupling, are perfectly compatible.
In comparison to a non-degenerate single-band model, we find that in the
degenerate two-band case the increase of the electron-phonon coupling, or
equivalently the decrease of the phonon frequency, is accompanied by
the closing of a gap in the excitation spectrum.

Notice that, even at quarter filling (one electron per site)
the antisymmetric charge mode remains gapless. In fact the
$8k_F$ Umklapp processes only couple to the symmetric charge mode,
which eventually acquire a gap, but do not affect the antisymmetric
charge sector. Therefore at filling 1/4 the difference between
the weak coupling regime (where the whole spectrum has a gap, as we
previously discussed) and the strong coupling regime (where we find one
gapless branch of excitations) is even more pronounced.

\subsection{Overall phase diagram of the degenerate two-band model}

We are now able to discuss the phase diagram of the
two-band model both at weak and strong electron-phonon coupling.
The phase diagram is qualitatively drawn in Fig.~3. We find
three different regions:
\begin{itemize}
\item[{\sl i)}] at weak coupling or high phonon frequency, the system
has dominant SS fluctuations (SS in Fig.~3);
\item[{\sl ii)}] upon increasing the coupling constant or decreasing
the phonon frequency, we find a cross-over to a region of dominant $4k_F$
CDW (see Fig.~3);
\item[{\sl iii)}] finally, further inside this region, the
gap in the antisymmetric charge sector closes and the model has
dominant inter-band $2k_F$ density waves
which can be also interpreted as orbital density waves ($2k_F$-ODW in Fig.~3).
\end{itemize}

The addition of the on-site repulsion (\ref{onsiterepulsion}) does
not modify qualitatively this phase diagram until the Hund
coupling $\Gamma$ is sufficiently large to make the triplet state
for two electrons favorable [see Eq.~(\ref{multiplet})].
If $\Gamma$ is not so large, we expect that the on-site
repulsion simply reduces
the region of stability of SS in favor of CDW. Moreover it also
suppresses the CDW components where the charge fluctuates between 0
and 2, and increases the components where it fluctuates between 0 and 1.
This implies:  1) that the $8k_F$-CDW component is enhanced against the
$4k_F$ one; and 2) that the $4k_F$ pseudo angular momentum density wave
component is enhanced against the $2k_F$ one.

\section{Conclusions}
In summary we have investigated how level degeneracy and dynamical
Jahn--Teller effect influence the
phonon-induced attraction in a simplified model of two degenerate bands
coupled to a doubly degenerate optical phonon. We have shown that a
new intersite attraction is generated to fourth order in the electron-phonon
coupling which is not present in a standard one band model. This
new process has many advantages with respect to the on-site polaronic
attraction, for instance it is robust to the presence of local repulsion
and it does not lead to phase separation~\cite{Manini,Maninibis} .
However its strength is much smaller, at least in the weak coupling limit we
have analyzed, than the on-site polaronic term, so that it is still an open
question whether it may play a relevant role as a mechanism for
phonon-induced superconductivity.

In one dimension the model can be analyzed in detail in the
whole range of the parameters. We have calculated the phase diagram which
displays quite new features as compared to the one band model.
Together with a standard cross-over from a superconducting region
to a charge-density wave regime as the electron-phonon coupling increases
(or equivalently the phonon frequency decreases), we
have found a new region at strong coupling where a kind of
modulated-orbital ordering is present. Such orderings have been already
proposed in
connection to systems with magnetic ions, but they were mainly ascribed to the
strong electronic correlations\cite{Khomskii}.
We have shown that modulated orbital-ordering may also arise from
purely vibrational mechanisms.

\section{Acknowledgments}
We are grateful to A. Parola, S. Sorella, G. Santoro and N. Manini
for discussions. EC Sponsorship under ERBCHRXCT 940438 and of
NATO CRG920828 are acknowledged.

\newpage

\begin{figure}
\caption{Hopping processes in the subspace of the molecular ground
state configurations. In parenthesis we write the values of the
quantum numbers $S_z$ and $J_z$. We recall that
the lowest energy configuration of a doubly occupied site is a singlet.}
\label{fig1}
\end{figure}

\begin{figure}
\caption{Qualitative phase diagram
of a non-degenerate phonon mode coupled to a
single band of non interacting electrons. $g$ is the dimensionless
electron-phonon coupling
constant, $\omega_0$ is the phonon frequency and $t$ the electron hopping
matrix element. SS identifies the superconducting region while
$2k_F$-CDW the regime where charge density wave with $2k_F$ oscillations
dominate.
The spin modes are gapped everywhere in the phase diagram.}
\label{fig2}
\end{figure}

\begin{figure}
\caption{Phase diagram for a two-dimensional phonon mode coupled
to a doubly degenerate band of electrons (lattice of E-e molecules).
In this case $4k_F$-CDW indicates dominant density waves with period
$2\pi/(4k_F)$, while $2k_F$-ODW means orbital density waves
with period $2\pi/(2k_F)$. In parenthesis we indicate the density wave
components whose amplitude increases as the on-site repulsion gets bigger.}
\label{fig3}
\end{figure}


\begin{references}
\bibitem{nota1} More precisely $\Delta H$ contains terms of order at least
$g^4$ which do not commute with the Hamiltonian
$H_0=\omega_0(\vec{r}\,^2 +\vec{p}\,^2)/2$ of the free two-dimensional
oscillator. In fact (\ref{EffHamiltonian:eqn}) contains terms of order
$g^4$ which however commute with $H_0$.
\end{references}
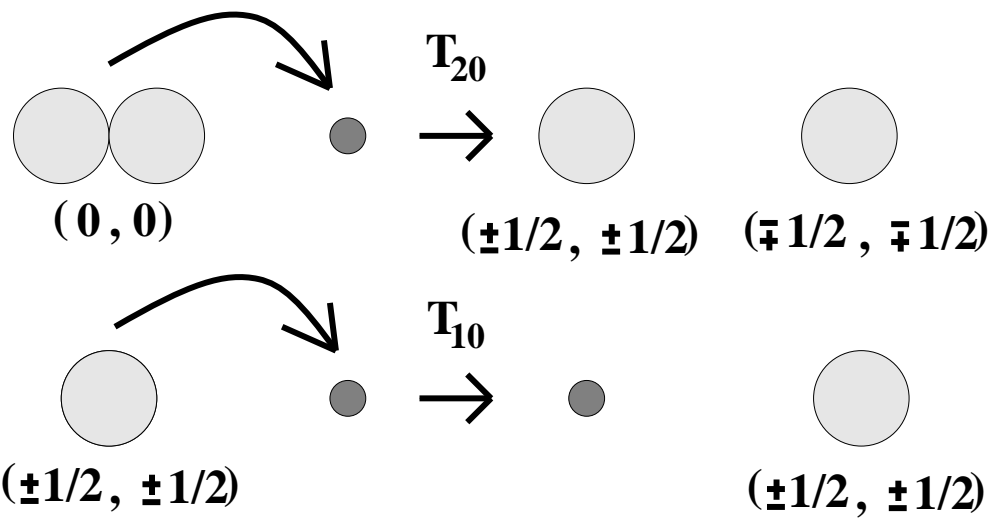
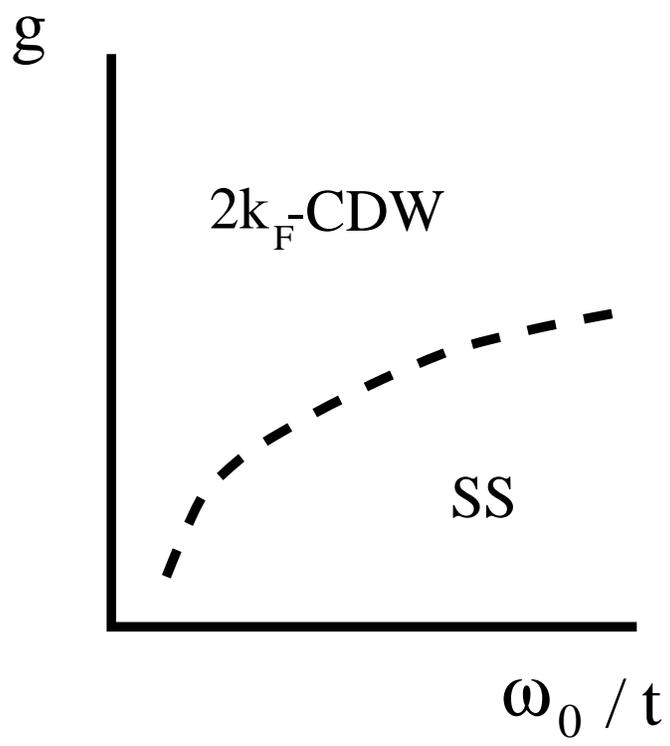
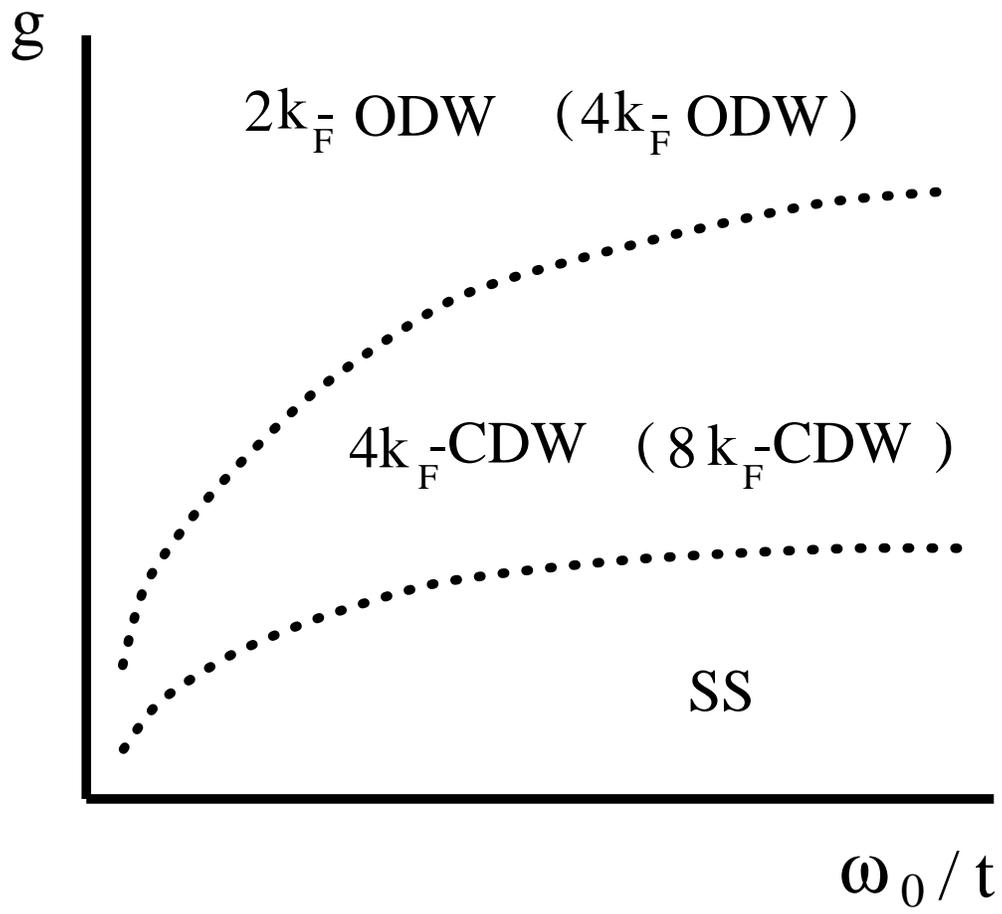
\end{document}